\documentclass{ws-p10x7}

\begin{document}

\title{New supersymmetric standard model with stable proton}

\author{Mayumi Aoki}

\address{Theory Group, KEK, Tsukuba, Ibaraki 305-0801, Japan \\
E-mail: mayumi.aoki@kek.jp}

\author{Noriyuki Oshimo}

\address{Department of Physics, Ochanomizu University \\
Otsuka 2-1-1, bunkyo-ku, Tokyo 112-8610, Japan\\
E-mail: oshimo@phys.ocha.ac.jp}  

\twocolumn[\maketitle\abstract{
We discuss a supersymmetric extension of the standard model with 
an extra U(1) gauge symmetry.  In this model, the proton stability is 
guaranteed by the gauge symmetry without invoking $R$ parity.  
The gauge symmetry breakdown automatically generates an effective 
$\mu$ term and large Majorana masses for right-handed neutrinos. 
The supersymmetry-soft-breaking terms for scalar fields could be 
universal at a very high energy scale, and the electroweak symmetry 
is broken through radiative corrections.  
}]

     The supersymmetric standard model (SSM) is considered 
to be one of the most 
plausible extensions of the standard model (SM).  
In particular, the minimal SSM (MSSM) is usually treated as the 
standard theory around the electroweak scale.  
However, this MSSM suffers one potentially 
serious problem.  
The proton may decay through baryon-number-violating  
couplings of dimension four, and thus its 
life time could become unacceptably short.  
In order to forbid those dangerous couplings, therefore, 
an {\it ad hoc} discrete symmetry is imposed on the model 
through $R$ parity.  
In the SM, on the other hand, the interactions which 
induce the proton decay are not allowed by gauge symmetry.  
The SSM would become more plausible, if the proton 
can be protected from decay more naturally.  

     In this report, we present a new SSM \cite{model1,model2}, 
in which the proton stability is guaranteed by an extra 
U(1) gauge symmetry, within the framework 
of a model coupled to $N=1$ supergravity.   
This model also provides natural explanations for the $\mu$ parameter 
and neutrino masses which are merely put by hand 
in the usual SSM.  

     Keeping the extension of the SM as minimal as possible, 
the particle contents of the model are taken as shown in Table 1.  
The extra U(1) gauge symmetry is denoted by U$'$(1),
for which the charges of superfields are expressed as 
$Q_Q$, $Q_{U^c}$, etc..  
The index $i$ $(=1,2,3)$ of the superfields for quarks and leptons stands 
for the generation, while the indices $j$ $(=1\cdots n_j )$ of $H_1$ and
$H_2$, $k$ $(=1\cdots n_k)$ of $S$, and $l$ $(=1\cdots n_l)$ of $K$ 
and $K^c$ are attached for possible multiplication.  

     In addition to the superfields of the MSSM, 
our model has SM gauge singlets $N^c$ and $S$ for, 
respectively, right-handed neutrinos and Higgs bosons to 
break the U$'$(1) symmetry.  
The superpotential is then required to contain the couplings 
$H_2LN^c$ and $SN^cN^c$ for giving non-vanishing but tiny
masses to the ordinary neutrinos.   
The $\mu$ term in the MSSM is replaced by $SH_1H_2$.  
New colored superfields $K$ and $K^c$ 
are incorporated to cancel a chiral anomaly.  
Their fermion components can receive masses from $SKK^c$.  

     The hypercharges of $K$ and $K^c$, the U$'$(1) charges 
of all the superfields, and the numbers $n_j$, $n_k$, and $n_l$ are 
determined by chiral and trace anomalies and necessary couplings 
of superfields.   Barring irrational hypercharges for $K$ and $K^c$, 
we obtain $Y_K=\pm \frac{1}{3}$ and $n_j=n_k=n_l=3$.  
The superfields with the same quantum numbers are all triplicated.  
The hypercharge $Y_K$ is either 1/3 or $-$1/3.  
However, the proton stability is satisfied only for $Y_K=1/3$.  
In this case, allowed couplings of dimension four 
are given by $H_1QD^c$, $H_2QU^c$, $H_1LE^c$, $H_2LN^c$,
$SH_1H_2$, $SN^cN^c$, and $SKK^c$.
The baryon number is conserved.  
The lowest dimension couplings with baryon-number violation 
are given by the D terms of $QQU^{c*}E^{c*}$,
$QQD^{c*}N^{c*}$, and $QU^{c*}D^{c*}L$, which are of dimension six.
The proton decay is adequately suppressed.  
On the other hand, for $Y_K=-1/3$, the particle contents of 
one generation can be 
embedded in the fundamental 27 representation of the E$_6$ group.  
As well known, the baryon and lepton numbers are not conserved in couplings 
of dimension four, such as $U^cD^cK^c$ and $LQK^c$, inducing 
a fast proton decay.  

     Requiring orthogonality between U(1) and U$'$(1) generators, 
the U$'$(1) charges of the superfields are determined up to 
a normalization factor.  
In Table 2, we show the U$'$(1) charges which are normalized 
to the U(1) charges.  

\begin{table}
\caption{Particle contents and their quantum numbers.  
$i=1,2,3$; $j=1,..,n_j$; $k=1,..,n_k$; $l=1,..,n_l$.  
        }\label{particles}
\begin{tabular}{l| c c c c}
   &  SU(3) & SU(2) & U(1) & U$'$(1) \\  
\hline
$Q^i$    & 3   & 2 & $\frac{1}{6}$  & $Q_Q$  \\ 
$U^{ci}$    & $3^*$ & 1 & $-\frac{2}{3}$ & $Q_{U^c}$  \\ 
$D^{ci}$    & $3^*$ & 1 & $\frac{1}{3}$  & $Q_{D^c}$  \\     
$L^i$    & 1   & 2 & $-\frac{1}{2}$ & $Q_L$  \\ 
$N^{ci}$    & 1   & 1 &      0         & $Q_{N^c}$ \\
$E^{ci}$    & 1   & 1 &      1         & $Q_{E^c}$  \\
$H_1^j$  & 1   & 2 & $-\frac{1}{2}$ & $Q_{H_1}$  \\
$H_2^j$  & 1   & 2 & $\frac{1}{2}$  & $Q_{H_2}$  \\
$S^k$    & 1   & 1 &      0         & $Q_S$   \\
$K^l$    & 3   & 1 & $Y_K$  & $Q_K$  \\
$K^{cl}$ & $3^*$ & 1 & $-Y_K$ & $Q_{K^c}$  \\ 
\end{tabular}
\end{table}
\begin{table}
\caption{U$'$(1) charges of the superfields.}
\label{charges}
\begin{tabular}{c c c c c c}
$Q_Q$ & $Q_{U^c}$ & $Q_{D^c}$ & $Q_L$ & $Q_{N^c}$ & $Q_{E^c}$      \\[1mm] 
\hline 
$\frac{1}{12}$ & $\frac{1}{12}$ & $\frac{7}{12}$ & $\frac{7}{12}$ & 
$-\frac{5}{12}$ & $\frac{1}{12}$ \\[1mm]
\hline
$Q_{H_1}$ & $Q_{H_2}$ & $Q_S$ & $Q_K$ & $Q_{K^c}$ &       \\[1mm] 
\hline
$-\frac{2}{3}$ & $-\frac{1}{6}$ & $\frac{5}{6}$ & $-\frac{2}{3}$ & 
$-\frac{1}{6}$ &  
\end{tabular}
\end{table}

     The superpotential is given by 
\begin{eqnarray}
W=&&\eta_u^{ijk}H_2^iQ^jU^{ck} + \eta_d^{ijk} H_1^iQ^jD^{ck} \nonumber \\
&&+ \eta_\nu^{ijk} H_2^iL^jN^{ck} + \eta_e^{ijk} H_1^iL^jE^{ck} \nonumber \\
&&+ \lambda_N^{ijk}S^iN^{cj}N^{ck} + \lambda_H^{ijk}S^iH_1^jH_2^k \nonumber \\
&& + \lambda_K^{ijk}S^iK^jK^{ck}, \nonumber
\end{eqnarray}
where all the couplings allowed by gauge symmetry and renormalizability 
are contained.  
The couplings are all cubic, and there is no mass parameter.  

     The model is coupled to $N=1$ supergravity, which 
breaks supersymmetry softly in the observable world.  
The Lagrangian contains, as well as supersymmetric terms, 
mass terms for scalar bosons and gauge 
fermions, and trilinear couplings for scalar bosons.  
In the ordinary scheme, the masses-squared and trilinear 
coupling constants for the scalar bosons have universal values 
$m_{3/2}^2$ and $A$, respectively, 
at the energy scale a little below the Planck mass.  
Hereafter, the scalar components of the superfields $H_1$,  
$H_2$, and $S$ are expressed by $\tilde H_1$,  
$\tilde H_2$, and $\tilde S$.  

     The parameter values of the model change according to 
the relevant energy scale.  
The mass-squared of $\tilde H_2^3$ 
receives large negative contributions through quantum
corrections, owing to a large coefficient $\eta_u$ of $H_2^3Q^3U^{c3}$ 
related to the top quark mass. As a result,  
this mass-squared becomes small around 
the electroweak scale, leading to non-vanishing vacuum 
expectation values (VEVs) for $\tilde H_1^3$ and $\tilde H_2^3$.  
The electroweak symmetry is broken through radiative corrections.  
On the other hand, for the first two generations, quantum corrections
to the masses-squared of $\tilde H_1^i$ or $\tilde H_2^i$ are not large, 
so that the VEVs of these scalar bosons vanish.  
If a coefficient $\lambda_K$ of $S^3K^3K^{c3}$ is large, the mass-squared of 
$\tilde S^3$ is also driven small.  A non-vanishing VEV is induced 
for $\tilde S^3$, and the U$'(1)$ symmetry is broken spontaneously.  

     The scalar potential is numerically analyzed at the electroweak 
scale to examine the parameter regions which give a vacuum 
consistent with experimental results.  
In particular, the U$'$(1) symmetry predicts a new neutral 
gauge boson $Z'$, for which stringent constraints are 
obtained on the mass and the mixing with the $Z$ boson of the SM.  
We assume that $\tilde H_1^3$, $\tilde H_2^3$, and $\tilde S^3$ 
of the third generation have non-vanishing VEVs, 
which are denoted by $v_1$, $v_2$, and $v_s$.  
It is shown that sizable regions are allowed for the mass-squared 
parameters $M_{H_1}^2$, $M_{H_2}^2$, and $M_S^2$ of the Higgs bosons.  
The coefficient $\lambda_H$ of $SH_1H_2$ should be around $0.1-0.4$.   
Owing to the constraints from the $Z'$ boson, 
$M_{H_1}^2$ is mostly larger than (1 TeV)$^2$.  
The value of $M_{H_2}^2$ is generally smaller than $M_{H_1}^2$
in magnitude, and $M_S^2$ is always negative.
The VEV of $\tilde S$ is larger than 1 TeV.  

     The term $\lambda_H SH_1H_2$ assumes the $\mu$ term of the SSM.    
The effective $\mu$ parameter is given by $\lambda_H v_s/\sqrt{2}$, 
which has an appropriate magnitude for the electroweak symmetry breaking.  
The terms $SN^cN^c$ induce large Majorana masses 
for the right-handed neutrinos.   
The ordinary neutrinos have non-vanishing masses 
approximately given by
$|\eta_\nu|^2v_2^2/2\sqrt{2}|\lambda_N|v_s$.
Taking for $\lambda_N\sim\lambda_H$, 
the neutrino masses become very small if  
the coupling constants $\eta_\nu$ for the neutrino Dirac masses are 
of the order of that for the electron $\eta_e$.  
A typical mass scale of squarks and sleptons is given by 
the universal value $m_{3/2}$ for the scalar boson masses, 
which is around $M_{H_1}$ and thus of order 1 TeV.  
Then, the smallness of the neutron and the electron 
electric dipole moments, which is another problem 
in the MSSM, can be explained.  

     In this model, the lightest Dirac fermion $\psi_K$ in the $K$ and $K^c$ 
system is stable, having both color and electric charges.  
In the early universe, after going out of thermal equilibrium,  
this fermion is bound to become a color-singlet particle.  
Since its mass is large, the decoupling occurs much 
earlier than for the up and down quarks.  
Therefore, the bound state is mainly formed by $\psi_K$ and 
its anti-particle.  
This is an electrically neutral meson, and eventually decays into 
lighter particles.   
The remnants for $\psi_K$ could exist in the present universe, and  
may be explored by non-accelerator experiments.    
However, its relic density depends on various uncertain factors, 
making a definite prediction difficult.

     The energy dependencies of the model parameters 
are quantitatively described by renormalization group equations.  
The coupling constant $\eta_u$ for the top quark mass should 
be around unity at the electroweak scale, which is obtained  
for $\eta_u>0.1$ at the high energy scale.  
The coupling constant $\lambda_K$ evolves similarly.  
If both $\eta_u$ and $\lambda_K$ are larger than 0.1 at the 
high energy scale, the mass-squared parameters  
$M_{H_2}^2$ and $M_S^2$ receive large quantum corrections.   
Taking the universal values as $m_{3/2}^2\sim$ (1 TeV)$^2$ 
and $A\sim 1$ for the mass-squared parameters and 
the trilinear coupling constants,
the values of $M_{H_2}^2$ and $M_S^2$ become small enough 
at the electroweak scale to induce 
SU(2)$\times$U(1)$\times$U$'$(1) gauge symmetry breaking.  
The quantum corrections to these parameters also become large, 
if the gauge fermion is heavy.   
On the other hand, the energy dependence of $M_{H_1}^2$ 
is weak and its value is not much different from the universal value.   
It is shown that there are reasonable parameter regions 
at the high energy scale, with the masses and trilinear 
coupling constants for scalar bosons being universal, 
which give parameter values at the electroweak scale 
consistent with phenomena.  
Unfortunately, however, the gauge coupling constants 
are not unified at an energy scale for possible grand unification,  
unless the particle contents of the model are modified.


\begin{thebibliography}{99}

\bibitem{model1}
M. Aoki and N. Oshimo, {\it Phys. Rev. Lett.} {\bf 84}, 5269 (2000).

\bibitem{model2}
M. Aoki and N. Oshimo, {\it Phys. Rev.} D {\bf 62}, 055013 (2000).

\end{thebibliography}
\end{document}